\documentclass[fleqn,twoside]{article}
\usepackage{espcrc2}
\usepackage{amssymb}
\usepackage{euscript}
\usepackage[T1]{fontenc}
\usepackage[latin1]{inputenc}
\usepackage{graphicx}
\usepackage{longtable}
\usepackage{pslatex}
\usepackage{subfigure}
\usepackage{epsfig}
\title{PHOTOS Monte Carlo and its theoretical accuracy       
}

\author{{Z. W\c{a}s}$^{a,b}$,
       P. Golonka$^a$,
       G. Nanava$^{c,d}$
        \\
{$^a$} CERN, 1211 Geneva 23, Switzerland \\
{$^b$} Institute of Nuclear Physics, PAN,
                   Krak\'ow, ul. Radzikowskiego 152, Poland\\
{$^c$} Universitaet Bonn, Physiklisches Institut, Nussallee 12, 53115 Bonn, Germany \\
{$^d$} On leave from IHEP, TSU, Tbilisi, Georgia%\\
}
\begin{document}

\begin{abstract}
           {Because of properties of QED, the bremsstrahlung corrections to decays
          of particles or resonances can be calculated, with a good precision, separately
	  from other effects. Thanks to the
	  widespread use of {\it event records} such calculations can be
	  embodied into a separate module of Monte Carlo
	  simulation chains, as used in High Energy Experiments of today.
	  The {\tt PHOTOS}
	  Monte Carlo program is used for this purpose since nearly 20 years now.
	  In the following talk let us review the main ideas and constraints which shaped
	  the program version of today and enabled it widespread use. Finally, we will 
          underline importance of aspects related to reliability of program results: 
          event record contents and implementation of channel specific matrix elements.

\noindent
{\bf 
    IFJPAN-IV-2008-1 \\ 
     May 2008}
}
\vskip 4 mm

\noindent
{ Presented by Z. W\c{a}s April 08 2008 on 
International Workshop on $e^+e^-$ collisions from $\Phi$ to $\Psi$,
                  \\
Laboratori Nazionali di Frascati, Italy
}

% #################
% ################## PoS format
% ##################
\vspace{1pc}
\end{abstract}

% typeset front matter (including abstract)
\maketitle

\section{\bf  Introduction}

In the construction of complex modular simulation systems, as the one used in High Energy
Physics, the question of dividing the system to functional parts is essential.
In the modular approach, the problem  can be divided into parts, and each part
can be addressed by different researchers or teams. However, such approach is only
possible if the mathematical structure of the problem have certain algebraic properties.

In practice, the module-based work model is often an idealization:
creating scientific software can not be separated from the research itself.
As a consequence, the architecture of programs needs to be modified at various
steps of projects development, for example to accommodate more precise models.

Another class of difficulties in the development of a simulation
segment arise from the constraints, imposed by other segments of the large project,
namely in the definition of data structures and interfaces.

Finally, the demands of the end-users, their expertise in handling simulation
blocks (which is often limited) and understanding the conceptual models used
are of critical importance.

These types of difficulties are quite universal for any complex, scientific problem.
In the following let us concentrate
on a relatively simple model (yet already quite sophisticated) for the simulation
of QED radiative corrections in decays, and the corresponding simulation package {\tt PHOTOS}
\cite{Barberio:1990ms,Barberio:1994qi,Golonka:2005pn,Golonka:2006tw}.
We believe that presented ``development drama'' can be of interest not only to the
readers interested in QED bremsstrahlung, but also in general case. In this respect
our presentation can be understood as a summary and invitation to reading \cite{PhDGolonka}.
With respect to previous \cite{Was:2007nf} longer version of the presentation let us point on the
particular property of the simulation of decays and in particular necessary in that step
radiative corrections. Functionality of such modules in some cases can be attributed 
to the physics of hard processes but in the other ones to the periferic ones, 
often to be used to measure spin states of the decaying particles. In the second case 
lower energy experiments provide invaluable framework necessary to evaluate systematic 
error for high energy applications. Finally such simulation modules need to match 
(directly or with the help of tests) the 
ones of low energy $e^+e^-$ experiments. 

Our contribution is organized as follows. We start with the presentation
of the {\tt PHOTOS} algorithm in section 2. We highlight the role of
the event record in program's construction. We will also specify those properties
of event record that are necessary for the program precision.

Section 3 is devoted to the description of recent years' developments in the {\tt PHOTOS}
algorithm, which lead to the improvement of its performance, but at the same time
introduced more strict constraint on the event record used as a data source.
We will present our efforts in preventing the complications for other users of data structures.

During the years of project's evolution
special techniques devoted to detect and resolve various types of difficulties related to
event record(s) were developed.
Section 4 will be devoted to their presentation.
Internal instrumentation utilities of {\tt PHOTOS}: debugging subroutine {\tt PHLUPA} and
kinematic rounding error correction subroutine {\tt PHCORK}  will be discusses.
Then, {\tt MC-TESTER} \cite{Golonka:2002rz} will be briefly presented.
Finally, we will justify the need to duplicate the event information which is originally
stored in a standard event record {\tt HEPEVT}.

Section 5 will be devoted to the discussion of the issues related to event record standardization.
The question of porting the algorithm to C++ - the process technically completed in 1999 -
\cite{MsCGolonka} will be covered in that section as well.
We will stress the important, yet often ignored, aspect of event record construction:
the structure must not only be convenient and flexible from the software engineer point of view,
its contents must also be clear, from the point of view of involved physics models, to the
end-users processing the data produced by Monte Carlo event generators.
This is particularly important if extensions of the standards are to be
not only proposed but also used.

The summary, Section 6, closes the paper.

\section{ Basic design of PHOTOS }

Already at an early step of preparation for the $\tau$-lepton polarization measurement
at LEP1 it became evident \cite{Boillot:1988re}, that bremsstrahlung  corrections in
$\tau$ decays are necessary for proper modeling of theoretical predictions
for measurement of the $Z$ boson couplings using $\tau$-polarization. For that purpose,
special routine {\tt RADCOR} was designed. With certain probability it was replacing
the decay products of $\tau$ simulated using {\tt TAUOLA} \cite{Jadach:1990mz} by the ones
with an extra photon added. This action was performed in a strict environment
of the explicit list of momentum four-vectors and in the rest-frame of $\tau$.
In fact, the parameter defining the actual decay mode of $\tau$ was also passed into {\tt RADCOR},
identifying the physical process.

The  $\tau$ decay products, modified in such way, were passed into the {\tt TAUOLA} interface
to {\tt KORALZ}
\cite{Jadach:1991ws}, which was the Monte Carlo program for $\tau$ pair production at LEP1 energies.

This awkward, yet useful, design was only tailored for a single application, and
was missing the documentation. The limits of its reliability were not explored at all.
Indeed, the fact that it could properly simulate the leading-log parts of the first-order
corrections was enough. The issues of dependence on the choice of the gauge were not risen,
nor were the question of the phase-space coverage.

However, this exercise provided an important observation: to simulate the dominant
part of QED corrections it is sufficient to search through the full event structure,
identify the branchings corresponding to elementary decays of some particles,
extract the information describing all decay products, and apply a routine such as
{\tt RADCOR}.
In the early 90's and late 80's the {\tt HEPEVT} event record was providing
the complete environment that could allow for extension in use of  {\tt RADCOR} routine.
The first version of {\tt PHOTOS} \cite{Barberio:1990ms} was
created. Its design relied heavily on the assumption that the {\tt HEPEVT} event record
hosts a tree-like structure and that all pointers to decay products (daughter pointers)
and origins (mother pointers) are defined and consistent.
At each decay splitting the
energy-momentum conservation was supposed to be fulfilled exactly, even if only with
a single-precision computer arithmetics level.

The {\tt PHOTOS} generator gained popularity, and in the documentation of its 1994 version
\cite{Barberio:1994qi} a multitude of tests were presented for physical processes
for which the theoretical predictions were available. The corrections for double
bremsstrahlung were added as well.
Further development was stalled due to limited interest in improvements,
 expressed by experimental
users.

\section{Toward high precision in PHOTOS }

Already in \cite{Barberio:1994qi} it was found, that if the algorithm for the single-photon
generation is properly iterated, the leading corrections of the double-photon emissions
can be incorporated as well. In  \cite{Golonka:2005pn} this iterative solution was extended to
multiple-photon emission and the tests for leptonic $Z$ decays have shown
that this solution reproduce the results for the final-state bremsstrahlung of
the {\tt KKMC} Monte Carlo program \cite{kkcpc:1999} with amazing precision.
Few years earlier, other tests and extensions important
for Higgs or $W$ decays were introduced also \cite{Andonov:2002mx}. The
good performance of the program was related to other improvements, introduced at that time and valid for all decays. For example
the implementation became exact in the soft-photon limit.
 These improvements, however,
are of limited importance from the point of view
of software organization of {\tt PHOTOS} interface with other packages.

Let us now concentrate on another class of corrections, discussed in
refs. \cite{Golonka:2006tw,Nanava:2006vv} and corresponding to
the implementation of the exact, first-order matrix-element kernel in
bremsstrahlung correction generation for $Z \to l^+ l^-$ and $ B \to K (\pi) K (\pi)$
decays. In the latter case, all possible combinations of charges and the replacement of
$\pi$ with $K$ were used.

Let us start with the presentation of the numerical results of the test for  $Z \to \mu^+ \mu^-$ decays taken from ref.~\cite{Golonka:2006tw}.

For that purpose we have to analyze the  plots giving largest discrepancies
between {\tt PHOTOS} (run with standard options) and {\tt KKMC} (run with second order matrix element and
exponentiation). The plots presenting the invariant mass of the $\mu^+\mu^-$ pair and the invariant mass
of the two hardest photons for the events with at least two photons of energies above 1 GeV in the rest frame of
the $Z$ are convenient for that purpose.
The rates for the event samples predicted by the two programs are given in the figure's caption.
The agreement between {\tt KKMC} and {\tt PHOTOS}
is better than 0.1 \% (if calculated with respect to the total $Z$ decay rate),
yet the differences are still visible from the results of simulations with $10^8$ events.
If, as in figure 2, the complete NLO kernel is activated in {\tt PHOTOS}, the differences get reduced by
about a factor of 50! This is rather  interesting result of ref. \cite{Golonka:2006tw}.

At this point it is important to realize what is the price to pay for such
improvement. Certainly, it is not the computer time  - it remains small and the samples
of order of $10^9$ could easily be simulated overnight ({\tt PHOTOS} is in fact significantly
faster than {\tt KKMC}). To answer this question one has to recall the formula for the final
Monte Carlo weight in {\tt PHOTOS}.

Let us write (separated from the phase-space Jacobians)
the explicit form  of the real-photon matrix element, as
used in the standard version of {\tt PHOTOS} (as published in \cite{Barberio:1990ms,Barberio:1994qi})
for the $e^{+}e^{-} \to Z^{0}/\gamma^{*} \to \mu^{+}\mu^{-} (\gamma)$ process:
{\small
\begin{eqnarray} &
X_{f}^{\mathrm{PHOTOS}}=\frac{Q'^{2}\alpha(1-\Delta)}{4\pi^{2}s}s^{2} \frac{1}{k'_{+}+k'_{-}}
(1+(1-x_{k})^{2})   \hskip 3 mm \Bigg\{  \nonumber \\ &
\frac{1}{k'_{-}} \bigg[
\frac{{d}\sigma_{B}}{d\Omega}\Big(s,\frac{s(1-\cos\Theta_{+})}{2},
\frac{s(1+\cos\Theta_{+})
}{2}\Big)\bigg]\frac{(1+\beta\cos\Theta_{\gamma})}{2}\;\;\; \nonumber\\ &
+
\frac{1}{k'_{+}} \bigg[
\frac{{d}\sigma_{B}}{d\Omega}\Big(s,\frac{s(1-\cos\Theta_{-})}{2},
\frac{s(1+\cos\Theta_{-})
}{2}\Big)\bigg]\frac{(1-\beta\cos\Theta_{\gamma})}{2}\Bigg\} \nonumber \\ &
\mathrm{where:}  \Theta_{+}=\angle(p_{+},q_{+}),\; \Theta_{-}=\angle(p_{-},q_{-}),
\; \nonumber\\
& \Theta_{\gamma}=\angle(\gamma,\mu^{-})\;  \textrm{is\, defined\,
in}\;(\mu^{+},\mu^{-})\textrm{-pair\, rest\, frame.} \nonumber
\label{X-fotos}
\end{eqnarray}
}
For its calculation (with respect to Born cross-section)
it is enough to know the four momenta of the $Z$ and its decay products.
In the presented formulae we follow the notations from
refs.~\cite{Golonka:2006tw,Berends:1982ie}.
This expression  is to be compared with the exact one, taken from
ref.~\cite{Berends:1982ie}:
{\small
\begin{eqnarray}
X_{f}=\frac{Q'^{2}\alpha(1-\Delta)}{4\pi^{2}s}s^{2} &
\Bigg\{\frac{1}{(k'_{+}+k'_{-})}\frac{1}{k'_{-}}\bigg[\frac{{d}\sigma_{B}
}{{d}\Omega}(s,t,u')+\frac{{d}\sigma_{B}}{{d}\Omega}(s,t',u
)\bigg]\nonumber \\
+ &
\frac{1}{(k'_{+}+k'_{-})}\frac{1}{k'_{+}}\bigg[\frac{{d}\sigma_{B}}{{d}
\Omega}(s,t,u')+\frac{{d}\sigma_{B}}{{d}\Omega}(s,t',u)\bigg
]\Bigg\}.\nonumber
\label{X-mustraal}
\end{eqnarray}
}
The resulting weight is rather simple, and reads:

{\small
\begin{eqnarray}
 WT_1=&  \frac{\frac{{d}\sigma_{B}
}{{d}\Omega}(s,t,u')+\frac{{d}\sigma_{B}}{{d}\Omega}(s,t',u
)}{\bigg[(1+(1-x_{k})^{2})
\frac{{d}\sigma_{B}}{d\Omega}\Big(s,\frac{s(1-\cos\Theta_{+})}{2},
\frac{s(1+\cos\Theta_{+})
}{2}\Big)\bigg]\frac{(1+\beta\cos\Theta_{\gamma})}{2}\; \big(1+ \frac{3}{4} \frac{\alpha}{\pi}\big)},    \hskip 5 cm    \nonumber \\
 WT_2=&   \frac{\frac{{d}\sigma_{B}}{{d}
\Omega}(s,t,u')+\frac{{d}\sigma_{B}}{{d}\Omega}(s,t',u)}{\bigg[(1+(1-x_{k})^{2})
\frac{{d}\sigma_{B}}{d\Omega}\Big(s,\frac{s(1-\cos\Theta_{-})}{2},
\frac{s(1+\cos\Theta_{-})
}{2}\Big)\bigg]\frac{(1-\beta\cos\Theta_{\gamma})}{2}\; \big(1+ \frac{3}{4} \frac{\alpha}{\pi}\big)}. \hskip 5 cm\nonumber
\label{wgt1}
\end{eqnarray}
}

For its calculation the numerical
value of the electroweak couplings of $Z$ to fermions, as well as information on the state
from which the $Z$ was produced is nonetheless necessary. This seemingly trivial requirement puts
new requirements on the event record: the details of the process of the $Z$ productions need to be coded
in the event record, then correctly deciphered by {\tt PHOTOS} to calculate the process-dependent
weight. From our experience this requirement of {\tt  PHOTOS} may be difficult to accept by
other users of event records. The authors of event generators often choose their own conventions
in encoding the details of hard process such as  $q \bar q \to ng Z/\gamma^*; Z/\gamma^* \to \mu^+ \mu^-$
into the event record.

The NLO solution for {\tt PHOTOS} would therefore be feasible with some universal, {\it standard} event record,
nonetheless difficult due to practical issues of interfacing. However, 
the NLO precision in {\tt PHOTOS} for today and tomorrow experiments
is most likely not required. For the time being the problem remain rather academic.
That is why we recommend the users to remain with the program version as distributed since
October 2005. Only the users interested in special tests and in particular incorporation of decay channel dependent electromagnetic formfactors are advocated to incorporate
further optional changes. In such a case even larger care on event record content is required.

In ref \cite{Nanava:2006vv}, we presented similar modifications in the {\tt PHOTOS} kernel
for the decay of $B$ mesons into a pair of scalars. The implementation of the exact (scalar-QED only) kernel brings
a minuscule
improvement in the agreement between {\tt PHOTOS} and the reference exact simulation of
{\tt SANC} \cite{Andonov:2004hi}.
In this case both: {\tt SANC} and {\tt PHOTOS} are used to simulate single photon emission
(There exists no reference simulation with which the multi-photon version of {\tt PHOTOS} could be
compared.).

The way to play with  form-factors, originating from phenomenological models, and fits
 to the experimental data is open now.

.

\section{ PHOTOS debugging tools}

During the years of {\tt PHOTOS} development, various software-related problems needed
to be faced by its authors. The majority of the workload (and actual lines of code)
needed to be devoted to the treatment of the data stored in a ``standard''
{\tt HEPEVT} event record.

In an ideal situation, if all rules of the {\tt HEPEVT} standard definition were respected, {\tt PHOTOS}
would easily identify the branching points (particle decays with charged products) in the decay tree,
extract the required data,
then eventually append the generated photons (if any) as additional decay products.
It would also traverse the decay tree to identify all possible places where QED
corrections might need to be generated. Initially, the algorithms employed in {\tt PHOTOS}
assumed that the data structure is consistent, with all pointers (to ``mothers'' and ``daughters'')
set up correctly. An acyclic tree of $1 \to n$ (or exceptionally $2 \to n$) processes could have easily
been navigated using standard algorithms.

However, the rigidness of the {\tt HEPEVT} standard, and the lack of possibility of extending it in a consistent
way forced the authors of event generators to overload the meaning of the elements of the {\tt HEPEVT}
data structure. Certainly, one could consider having bi-directional relations (i.e. mothers pointing
to daughters, and daughters pointing to mothers" as redundancy, and the place where additional information
(such as spin or colour flow) may be stored instead. The meaning of the pointers become generator-specific
and the navigation in such data structure could not be performed by a generic algorithm. Pandora's box
of event-record problems has been opened, hurting mainly the coordinators of the large experimental simulation
chains.

The pointers in the {\tt HEPEVT} structure were not the only element that became non-standard. Due to evolving
needs of physics models (such as bigger number of particles being simulated, or precision), {\tt HEPEVT} data
structure has been modified
to store single- or double-precision data, with various array sizes. Dubious matching of the {\tt HEPEVT}
layout between simulation blocks became yet more complicated.

To alleviate the problem with varying precision and layout of {\tt HEPEVT}, {\tt PHOTOS} has been equipped with a set
of debugging and data-interpretation facilities. Firstly, it was modified to work on a local copy of the
event record (the layout of which followed the "well-behaved" {\tt HEPEVT} standard), and have a set of functions
that would transfer the data between whatever external variant of {\tt HEPEVT} data structure, and the internal storage.
Secondly, a set of sanity-checking and pointer-reconstructing procedures were applied during the transfer
of the data between the external event record and the internal one.
Finally, a debugging function {\tt PHLUPA} was provided. It prints out the data
as interpreted/modified  by {\tt PHOTOS} routines, at different steps of event
construction.

Because of the specifics of the {\tt PHOTOS} algorithm, namely massive search and modification of the complete event
tree, {\tt PHOTOS} itself has became a debugging tool for large simulation chains in the experimental collaborations.
Strengthened with its debugging tools, it helped to identify many problems related to event grammatic.

This class of problems will remain with C++ applications as well. 
The origin of the difficulty is physics. On theoretical side authors of main 
Monte Carlo programs explore different options on storing  information 
on the process generation, on the other, experimental users prefer to rely only on 
the information, which can be obtained from the measurement.

To be able to explore, in such conditions, the potential of the {\tt PHOTOS} algorithm, and to make the debugging of  the
event record data easier, a  tool:
{\tt MC-TESTER} \cite{Golonka:2002rz} originally developed for tests of {\tt TAUOLA} was adopted. It performs comparison tests of distributions of invariant masses produced by two,
or more, (versions of) event generators. In semi-automatic way (thus eliminating the risk of accidental
programmatic error) it extract the data from event records filled by an event generators: it identifies
the decay modes of a given particle, and for every mode it builds the histograms of invariant masses
of all combinations of decay products.

\section{Challenges of event record}

On reading the paper, one could get an impression that the communication between the modules
of the simulation tree is a challenging, yet standard,
goal, which could be realized in any modern, or even not so modern software
environment. Our discussion in the previous chapters pointed to possible
constraints and requirements in the organization of such data structure
imposed by relatively modest
application {\tt PHOTOS}, if its precision requirements would need to be increased beyond certain
level.

One can ask the question why the C++ implementation of {\tt PHOTOS} did not meet so far
as much attention as its seemingly obsolete FORTRAN version. On a first
sight the answer is simple: there was until recently no commonly agreed standard
for the event record data structure in C++ accepted by the dominant  part of
the community. Recently, it seems that the HepMC \cite{Dobbs:2001ck} structure is gaining popularity
in the LHC applications.

From the past experience of {\tt HEPEVT} standard, one could postulate that a viable event record
could be seen as a system of parallel trees,  the nodes of each being bound with
correspondence relations spanning across the layers.
Each individual tree should be easy to
investigate or modify by program such as {\tt PHOTOS}, or the detector simulation
software.

For {\tt MC-TESTER} to be effective in comparing the results generated by various
Monte Carlo simulators, it was essential to provide access to various standards (or:
flavors) of event record data structures. Similarly to {\tt PHOTOS}, {\tt MC-TESTER}
performs exhaustive search and data extraction from event record, it however
doesn't need to modify the contents of the event record. Typical data that needed
to be extracted was the mother-daughter relationships (including finding out
the non-decaying final-state particles in the cascade decay), and determining
the properties (four-momenta, and type) for involved particles. To separate {\tt MC-TESTER}
from problems related to event record processing, the {\tt HEPEventLib} abstraction layer
was created 

At the technical layer, {\tt HEPEventLib} does nothing more than interpretation of
the data stored in various event record standards, and providing these data
in a consistent form to the main program, hence hiding all dependencies
and data-translation operations. The data is provided by abstract object
representing a particle, a list of particles and an event.
As no modification is performed on the structure of the event record, the
properties visible in a "particle" object may be mapped directly to the
corresponding data in the underlying event record. Thanks to that feature,
particle's properties (such as four-momenta, but not the mother-daughter
attributions)
might even be modified from within {\tt HEPEventLib}'s abstract view, and the changes
would be propagated to the actual event record in a consistent way.

At the bottom of the {\tt HEPEventLib} there are implementations of the {\tt HEPEventLib}
abstraction to concrete event record types, of FORTRAN or C++ applications.
New "backends" for any (future)
event record may be implemented as needed - {\tt MC-TESTER} will automatically
profit from the new standard with no need of adding a single line of code in it.

\section{ Summary}

In the present talk we have reviewed the basic properties of theoretical (QED) and
software environment which is at foundation of design and performance of {\tt PHOTOS} Monte
Carlo program for simulation of QED bremsstrahlung corrections in decays of resonances
and particles encoded in different type of event records used in simulations of
High Energy physics. Special emphasis was on the points important for evaluation of 
systematic errors.

\centerline{\bf Acknowledgments}

This work is partly supported by  EU Marie Curie Research Training Network grant under the contract No. MRTN-CT-2006-0355505 and by the EU 6th Framework Programme under the contract No. MRTN-CT-2006-035482 (FLAVIAnet network).

%\bibliographystyle{JHEP}
%\addcontentsline{toc}{section}{\refname}\bibliography{ACAT07}
\providecommand{\href}[2]{#2}\begingroup\raggedright\endgroup

\end{document}